\documentclass[12pt]{article}
\usepackage{amsmath,amsbsy,amssymb,graphics}
\usepackage{graphicx,epsfig}
\def\beq{\begin{equation}}
\def\eeq{\end{equation}}
\def\be{\begin{equation}}
\def\ee{\end{equation}}
\def\bea{\begin{eqnarray}}
\def\eea{\end{eqnarray}}
\def\nnb{\nonumber}

\newcommand{\gsim}{\lower.7ex\hbox{$\;\stackrel{\textstyle>}{\sim}\;$}}
\newcommand{\lsim}{\lower.7ex\hbox{$\;\stackrel{\textstyle<}{\sim}\;$}}

\begin{document}

\begin{center}
 \vspace{0.2cm}
  {\large \bf Precise Formulation of Neutrino Oscillation in the Earth}

\vspace{0.6cm}
{\large \bf Wei Liao }

\vspace{0.3cm} { Institute of Modern Physics
\footnote{Previously Institute of Theoretical Physics} \\
 East China University of Science and Technology \\
 P.O. Box 532, 130 Meilong Road, Shanghai 200237, P.R. China}
\end{center}
\begin{abstract}
\vskip 0.2cm

\end{abstract}
 We give a perturbation theory of neutrino oscillation in the Earth. The
 perturbation theory is valid for neutrinos with energy $E \gsim 0.5$ GeV.
 It is formulated using trajectory dependent average potential. Non-adiabatic
 contributions are included as the first order effects in the perturbation
 theory. We analyze neutrino oscillation with standard matter effect and
 with non-standard matter effect. In a three flavor analysis we show that
 the perturbation theory gives a precise description of neutrino conversion
 in the Earth. Effect of the Earth matter is substantially simplified in
 this formulation.


\section{Introduction}\label{sec1}

One of the major concerns in neutrino oscillation experiments is the
effect of Earth matter in neutrino flavor
conversion\cite{msw1,msw2}. Because of the complicated matter
density profile in the Earth it is a challenge to find a simple
formula which can describe precisely neutrino conversion in the
Earth. We rely a lot on numerical computation which does not offer
us enough insights. Few years ago, an analytic and precise
formulation was obtained for solar neutrinos, i.e. for $E \lsim 30$
MeV \cite{hls,other}. It is the purpose of the present paper to find
a good formulation for higher energy neutrinos, i.e. for $E \gsim
0.5$ GeV. Previous works on neutrinos of this energy range include
\cite{some1,ak,ps,cer,fhl,ms,bo,ams2,others}.

It will be shown that the approach using trajectory dependent
average potential is very attractive. This approach was used in
\cite{ls} in discussing solar neutrinos. It was used
in \cite{ams1} in a $2\nu$ analysis for higher energy neutrinos ($E \gsim
0.5$ GeV). In the present paper we present a perturbation theory
for three flavors of neutrinos. The perturbation theory uses trajectory dependent average
potential. We do perturbative expansion using the deviation of potential
around the average. We analyze the perturbation theory using the density profile
of the Preliminary Earth Model(PREM) \cite{PREM}.
We study the perturbation
theory for the case of standard matter effect
and the case of non-standard matter effect with non-standard
Neutral Current (NC) interactions.
We show that the perturbation theory works very well for both
cases. The theory is valid for neutrinos with $E \gsim 0.5$ GeV.

The perturbation theory presented is very useful for long baseline
neutrino experiments \cite{K2K,Minos,T2K,opera,NoVa}. It says for a
fixed baseline ( $\lsim 6000$ km) the standard matter effect is a
one parameter fit. Non-standard matter effect is also greatly
simplified. The perturbation theory is of general interests to other
types of neutrino sources, e.g. for studying atmospheric neutrinos,
cosmic neutrinos from the galactic or extra-galactic sources. The
present paper is organized as follows. In section \ref{sec2} we
present the perturbation theory of three flavor of neutrinos. In
section \ref{sec3} we study neutrino oscillation in the case of
standard matter effect. In section \ref{sec4} we extend the
discussion on neutrino oscillation to the case of non-standard
matter effect with non-standard neutrino NC interaction. We discuss
and summarize in section \ref{sec5}.

\section{Formulation of $\nu$ oscillation in the Earth}\label{sec2}

 We consider oscillation of three flavors of neutrinos:
 $\psi=(\nu_e, \nu_\mu,\nu_\tau)$.  The evolution equation is
  \bea
 & i \frac{d}{d x} \psi(x) = H(x) \psi(x), ~~ H(x) = H_0 + V(x)
 \label{evol1}\\
 & H_0 =\frac{1}{2 E} U ~ \textrm{diag}\{0, \Delta m^2_{21}, \Delta m^2_{31}\}~ U^\dagger,
  \label{evol1b}
 \eea
 where $V(x)$, a $3\times 3$ matrix, is the potential term
 accounting for the matter effect. $V(x)$ takes different form if
 neutrino interaction is different. $U$ is the $3\times 3$
 neutrino mixing matrix in vacuum. $U$ is parameterized using
 standard parameters $\theta_{12}$, $\theta_{13}$, $\theta_{23}$ and
 $\delta_{13}$, the CP violating phase.

 In solving the problem of neutrino oscillation we introduce the average potential term
 ${\bar V}$:
 \bea
  ~~{\bar V}=\frac{1}{L} \int^L_0 ~d x~ V(x),
  \label{avepoten}
  \eea
 where $L$ is the length of neutrino trajectory in the Earth.
 Note that ${\bar V}$ depends on the trajectory of neutrino in the
 Earth. It is averaged over potentials along the trajectory and is not
 averaged over potentials of all points in the Earth. Using ${\bar V}$ we
 introduce ${\bar H}$ and the mixing matrix $U_m$ in matter which
 diagonalizes ${\bar H}$:
 \bea
 & {\bar H}=H_0+{\bar V}, \label{hamil1} \\
 & {\bar H} U_m = U_m \frac{1}{2 E} \Delta, ~~
   \Delta=\textrm{diag}\{\Delta^1, \Delta^2, \Delta^3\}.
   \label{defU}
 \eea
 $\frac{1}{2 E} \Delta^i(i=1,2,3)$ are three eigenvalues of ${\bar H}$.
 $\Delta^1 < \Delta^2 < \Delta^3$ is satisfied in our convention.
 $U_m$ is parameterized using parameters
 $\theta_{12}^m$, $\theta_{13}^m$, $\theta_{23}^m$ and the CP violating
 phase $\delta_{13}^m$. Note that we work in the convention of Hamiltonian
 introduced in (\ref{evol1b}) and (\ref{hamil1}). In typical cases
 where $3\times 3$ evolution problem can be reduced to $2\times 2$
 evolution problem, e.g. when hierarchy in eigenvalues are present,
 eigenvalues obtained in our convention can be related to
 eigenvalues solved in standard $2\times 2$ problem by shifting the
 phase of neutrinos and making the reduced hamiltonian traceless.

 We are ready to solve the evolution problem in (\ref{evol1}).
  Note that $H(x)$ can be re-written as
 \bea
  H(x)={\bar H} +\delta V(x), ~~ \delta V(x)=V(x)-{\bar V}.
  \label{hamil2}
  \eea
  We first solve the evolution by ${\bar H}$ and obtain
  the contribution of $\delta V$ using perturbation in $\delta V$.
  Keeping result of first order in $\delta V$ we obtain
 \bea
  \psi(L)&&=M(L) \psi(0), \label{evol2} \\
  M(L) &&=U_m ~e^{- i \frac{\Delta}{2 E} L}(1- i C) ~U^\dagger_m,
  \label{evol2b}
  \eea
 where $C$ is a $3\times 3$ matrix accounting for the non-adiabatic
 transition:
 \bea
 C=\int^L_0 dx ~e^{i \frac{\Delta}{2 E} x} U^\dagger_m \delta V(x) U_m
  e^{-i \frac{\Delta(x)}{2 E} x} \label{evol3}.
 \eea
 It is clear that $C^\dagger =C$ holds. One can see that
 \bea
 C_{jj} &&=\int^L_0 dx ~(U^\dagger_m \delta V(x) U_m)_{jj}=0, \label{C1}
 \\
 C_{jk} &&=\int^L_0 dx ~e^{i \frac{\Delta^j-\Delta^k}{2 E} x} (U^\dagger_m
 \delta V(x) U_m)_{jk}, ~~j\neq k . \label{C2}
 \eea
 Eq. (\ref{C1}) is guaranteed by Eq. (\ref{avepoten}).
 $|C_{jk}| \ll 1 (j\neq k)$ should be satisfied if this is a good perturbation
 theory. One of the virtues of this perturbation theory is that Eq. (\ref{C1})
 guarantees that the oscillation phase is correctly reproduced.

 In the next two sections we will discuss in detail that we indeed do expansion
 in small quantities when computing $C_{jk}(j\neq k)$ and
 correction of the second order ${\cal O}(C^2)$ is
 further suppressed by these small quantities. Hence we can convince ourselves
 that we are dealing with a perturbation theory. Before making detailed discussions
 on it we note that it is a perturbation partly
 because density changes mildly in the mantle or in the core of the Earth.
 In the mantle or in the core $||\delta V||/||V|| \lsim 0.3$.
 This helps in improving the perturbation theory.

  When $L > 10690$ km neutrinos cross the core of the Earth.
  Large density jump between the core and the mantle can cause problem
  to the simplest version of the perturbation theory.
  To improve the approximation we use average potentials for parts of
  trajectory in the core and in the mantle separately.
  So the evolution matrix can be written as
   \bea
   M=M_3 M_2 M_1, \label{CoreMantle}
   \eea
 where $M_2$ is the evolution matrix in the core and $M_{1,3}$
 are evolution matrices in the mantle. They are
 \bea
 M_i &&=U_{mi} ~e^{- i \frac{\Delta_i}{2 E} (L_i-L_{i-1})}(1- i C_i) ~U^\dagger_{mi},
 ~~i=1,2,3 \label{evol2c}
 \eea
 where $L_0=0$ and $L_3=L$. $x=L_1$ is the point where neutrinos cross from
 the mantle to the core. $x=L_2$ is the point where neutrinos come out of
 the core to the mantle. $\Delta_i$ and $U_{mi}$ are the engenvalues and
 mixing matrix in the region $x \in [L_{i-1}, L_i]$. They are computed
 using average potential
 \bea
 {\bar V}_i= \frac{1}{L_i-L_{i-1}} \int^{L_i}_{L_{i-1}} ~dx ~V(x). \label{avepotenb}
 \eea
 $C_i$ is computed using Eq. (\ref{evol3}) in the region $x \in [L_{i-1}, L_i]$.
 Because of the approximate symmetric density profile of the Earth we have
 \bea
 {\bar V}_3 \approx {\bar V}_1. \nnb
 \eea
 This strategy to treat evolution of core crossing neutrinos was also noticed
 in Ref. \cite{ams1}. In the present paper when we do computation using the perturbation
 theory we always use Eq. (\ref{CoreMantle}) for core crossing trajectories
 which have $L > 10690$ km and use Eq. (\ref{evol2b}) for trajectories
 crossing only mantle which have $L < 10690$ km.

 For anti-neutrinos the hamiltonian is
 \bea
 H^\prime(x) = H_0^T-V^T(x), \label{Hamil-anti}
 \eea
 where $V^T(x)$ is the transpose of the potential term $V(x)$ introduced in
 Eq. (\ref{evol1}). $H_0^T$ is the transpose of $H_0$. The problem
 of neutrino evolution can be solved using average potential, similar to the
 program used for neutrinos.

 The probability of neutrino oscillation is introduced using the
 evolution matrix:
 \bea
 P(\nu_k \to \nu_l) =| M_{lk}(x=L)|^2, ~~~
 P({\bar \nu}_k \to {\bar \nu}_l) = |{\bar M}_{lk}(x=L)|^2, \label{prob}
 \eea
 where ${\bar M}_{kl}$ is the evolution matrix of anti-neutrinos
 ${\bar \psi}=\textrm{diag}\{ {\bar \nu}_e, {\bar \nu}_\mu, {\bar \nu}_\tau \}$:
 \bea
 {\bar \psi}(L)= {\bar M}(L) {\bar \psi}(0). \nnb
 \eea
 Using the oscillation probability we can define observables of CP violation
 and of time reversal asymmetry. For example we define
 \bea
 A_{CP} && =\frac{P(\nu_\mu \to \nu_\tau)-P({\bar \nu}_\mu \to {\bar \nu}_\tau)}
 {P(\nu_\mu \to \nu_\tau)+P({\bar \nu}_\mu \to {\bar \nu}_\tau)}, \label{ACP} \\
 A_T &&= \frac{P(\nu_e \to \nu_\mu)-P( \nu_\mu \to \nu_e)}
 {P(\nu_e \to \nu_\mu)+P(\nu_\mu \to \nu_e)}, \label{AT}
 \eea
 One can see in Eqs. (\ref{evol1}) and (\ref{Hamil-anti}) that
 $A_{CP}$ does not contain pure information of fundamental CP violating
 parameter. Because the medium contains only matter but not
 anti-matter $A_{CP}$ is not zero even if CP violating phase
 $\delta_{13}$ in $U$ is zero.
 It is clear that powerful electron neutrino beam is needed to observe
 the time reversal asymmetry. Intense electron neutrino beam is probably
 available using muon storage technology in future experiments.

\section{Oscillation with standard matter effect}\label{sec3}

  \begin{figure}
\begin{flushleft}
\includegraphics[height=6.cm,width=8cm]{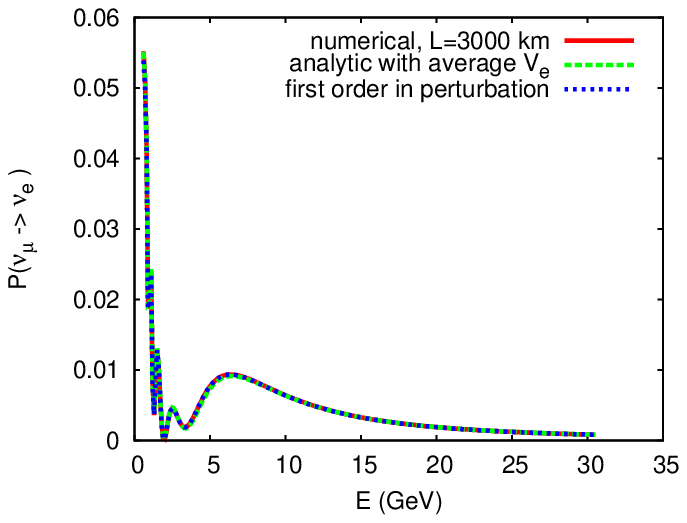}
\end{flushleft}
\begin{flushright}
\vskip -6.5cm
\includegraphics[height=6.cm,width=8cm]{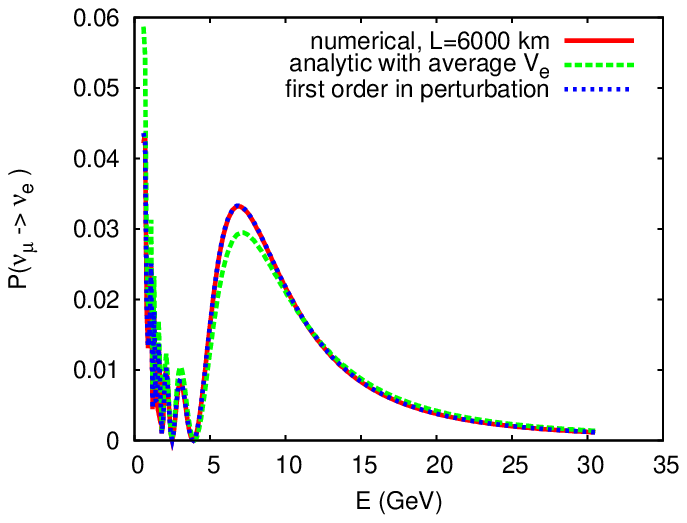}
\end{flushright}
\vskip 0.0cm \caption{\small $P(\nu_\mu \to \nu_e)$ versus energy
 for the case with standard matter effect.
  Left panel is for $L=3000$ km and
 right panel is for $L=6000$ km. $\Delta m^2_{21}=8. \times 10^{-5}$ eV$^2$,
 $\Delta m^2_{32}= 3. \times 10^{-3}$ eV$^2$. $\sin^2 2\theta_{23}=1$,
 $\tan^2 \theta_{12}=0.41$, $\sin^2 2\theta_{13}=0.01$, $\delta_{13}=\pi/6$.
 PREM density profile is used for computation in this figure and
 all remaining figures in this paper.}
 \label{PVsE}
 \end{figure}

  \begin{figure}
\begin{flushleft}
\includegraphics[height=6.cm,width=8cm]{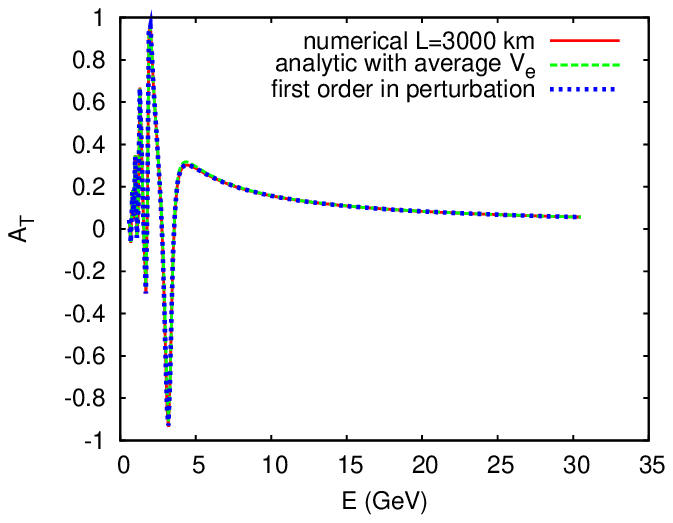}
\end{flushleft}
\begin{flushright}
\vskip -6.5cm
\includegraphics[height=6.cm,width=8cm]{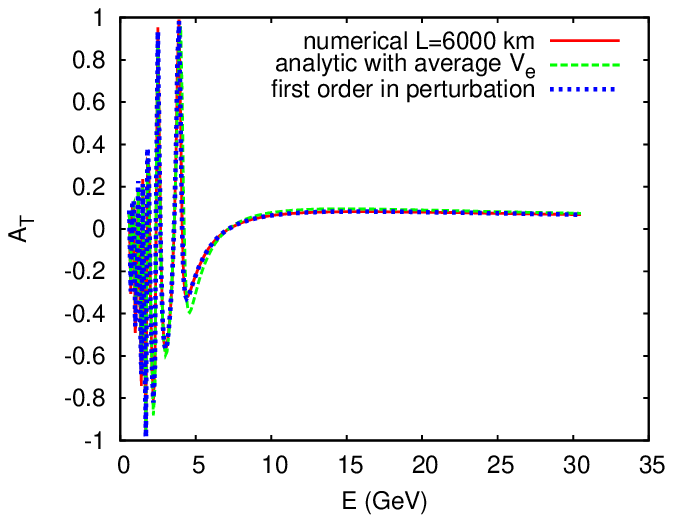}
\end{flushright}
\vskip 0.0cm \caption{\small $A_T$ versus energy
 for the case with standard matter effect.
 Left panel is for $L=3000$ km and
 right panel is for $L=6000$ km. Parameters are the same as in Fig. \ref{PVsE}.}
 \label{TsVsE}
\end{figure}

 In this section we consider oscillation of neutrinos with standard
 matter effect. In this case we express the potential, mixing matrix, etc
 as
 \bea
 V(x)=V_s(x),~~{\bar V}={\bar V}_s,~~H=H_s, ~~{\bar H}={\bar H}_s,~~U_m=U_s,~~\Delta=\Delta_s .
 \label{label1}
 \eea

 The potential term of standard matter effect is
 \bea
 ~~V_s(x)=\textrm{diag}\{V_e(x), 0,0\}.
 \eea
 $V_e=\sqrt{2} G_F N_e$ is the potential in matter. $N_e$ is the electron
 number density in matter. $G_F$ is the Fermi constant. Effect of standard
 NC interaction is universal for three types of neutrinos and is neglected
 for neutrino oscillation in Earth matter. ${\bar V}_s$, $H_s$,
 ${\bar H}_s$, $U_s$, $\Delta_s$ are defined using Eqs. (\ref{avepoten}),
 (\ref{evol1}), (\ref{evol1b}), (\ref{hamil1}) and (\ref{defU}).
 $C_{jk}$ is obtained as follows:
 \bea
 C_{jk}  = \int^L_0 dx ~e^{i \frac{\Delta^j_s-\Delta^k_s}{2 E} x} ~(U_s)^*_{ej}
  (U_s)_{ek} ~\delta V_e(x), ~~j\neq k, \label{C3}
 \eea
 where
 \bea
 \delta V_e(x)=V_e(x)-{\bar V}_e,
 ~~{\bar V}_e=\frac{1}{L} \int^L_0 ~dx~ V_e(x). \label{aveV}
 \eea

For trajectories crossing mantle only the evolution matrix is
 \bea
   M_s(L) &&=U_s e^{- i \frac{\Delta_s}{2 E} L}(1- i C) U^\dagger_s.
   \label{evolS}
 \eea
 For core crossing trajectories ($L > 10690$ km)
 the evolution matrix is generalized using Eq. (\ref{CoreMantle}).
 \bea
 M_s(L)= M_{s3} M_{s2} M_{s1}. \label{evolSb}
 \eea
 $M_{s2}$ and $M_{s1,s2}$ are evolution matrices in
 the core and mantle separately. They are computed
 to the first order in $\delta V$ and using average
 potentials in the core and the mantle separately, as explained
 in section \ref{sec2}.

 For comparison we also show the results computed using analytic
 formula:
 \bea
 M_{As}(L)=U_s ~e^{- i \frac{\Delta_s}{2 E} L} ~U^\dagger_s,
 \label{AnalyticA}
 \eea
 where non-adiabatic contribution is not included. This formula
 is not generalized for core crossing trajectories as done in
 Eq. (\ref{evolSb}).

 In Fig. \ref{PVsE} we plot $P(\nu_\mu \to \nu_e)$ versus energy. We show
 results of numerical computation, results computed using Eq. (\ref{evolS})
 and the result computed using analytic formula Eq. (\ref{AnalyticA}).
 The lines labeled with "analytic with average $V_e$" are computed
 using Eq. (\ref{AnalyticA}). The lines labeled with "first order
 in perturbation" are computed using Eq. (\ref{evolS}). They are
 plots for $L=3000$ km and $6000$ km. One can see that results computed
 using the perturbation theory reproduce precisely the
 phase and the magnitude of the oscillation pattern.
 One can see that for these two baselines
 the analytic results computed using Eq. (\ref{AnalyticA}) give
 a quite good approximation to the oscillation pattern. Actually one
 can not see difference in the left panel ($L=3000$ km). In the
 right panel ($L=6000$ km) one can see that oscillation phase is
 correctly reproduced by the analytic result but the magnitude around
 peaks is not precisely reproduced. There are some small differences.

 In the right panel height of the first peak of the right side is given
 by $\sin^2 2\theta^m_{13}$. One can see that it is much larger
 than the magnitude given by the vacuum value $\sin^2 2\theta_{13}$.
 This is because the position of the first peak
 is close to the region of $1-3$ MSW resonance which has energy range
 $E \sim 7-10$ GeV. Magnitudes of the second and third peaks are much smaller.
 This is because the second and third peaks are
 away from the resonance region and the vacuum value is dominant.
 In Fig. \ref{TsVsE} we show plots for the time
 reversal asymmetry $A_T$ versus energy. Again in the left panel the analytic
 result using Eq. (\ref{AnalyticA}) is precise. In the right panel there are
 some small differences between numerical and analytic results.
 Results computed using the perturbation theory, i.e. using Eq. (\ref{evolS}),
 are in remarkable agreement with the numerical results.

   \begin{figure}
\begin{flushleft}
\includegraphics[height=6.cm,width=8cm]{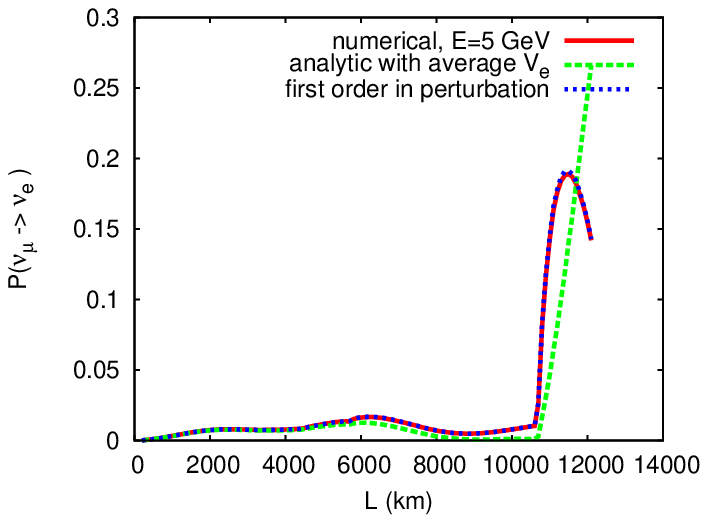}
\end{flushleft}
\begin{flushright}
\vskip -6.5cm
\includegraphics[height=6.cm,width=8cm]{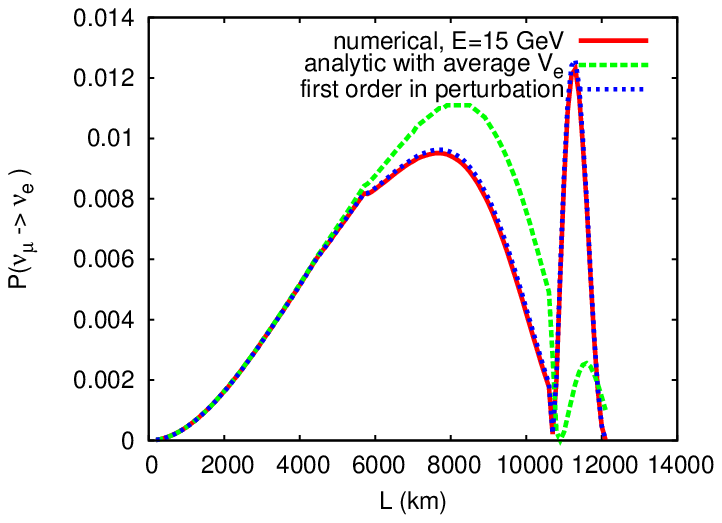}
\end{flushright}
\vskip 0.0cm \caption{\small $P(\nu_\mu \to \nu_e)$ versus L the length of
 baseline for the case with standard matter effect. Left panel is for $E=5$ GeV and
 right panel is for $E=15$ GeV. Neutrino parameters are the same as in Fig. \ref{PVsE}.}
 \label{PVsLA}
\end{figure}

  \begin{figure}
\begin{flushleft}
\includegraphics[height=6.cm,width=8cm]{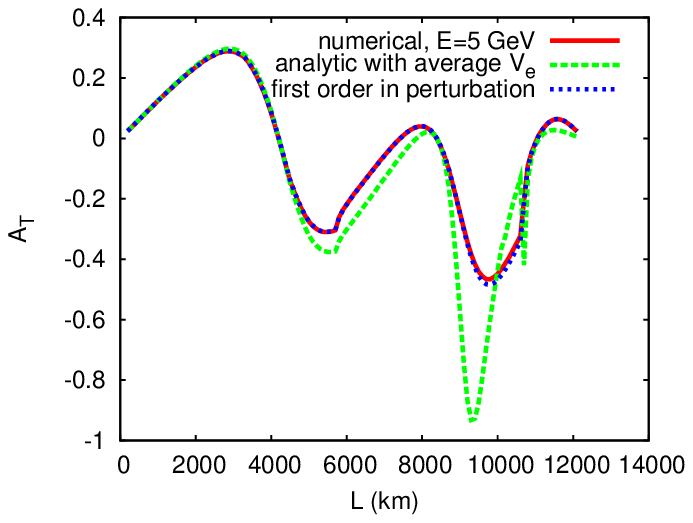}
\end{flushleft}
\begin{flushright}
\vskip -6.5cm
\includegraphics[height=6.cm,width=8cm]{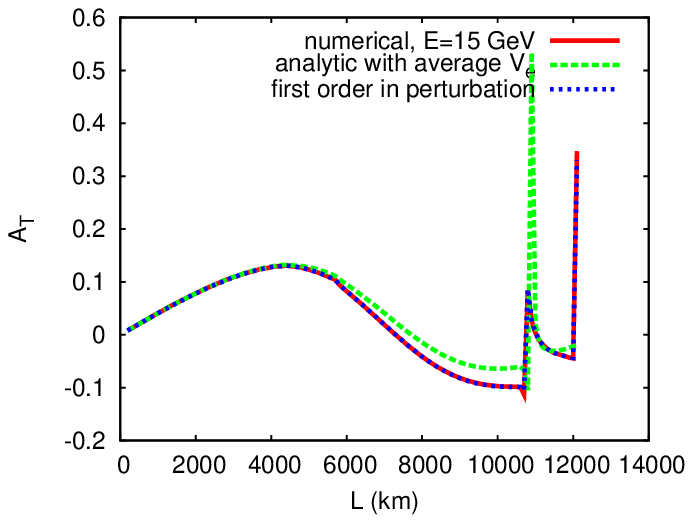}
\end{flushright}
\vskip 0.0cm \caption{\small $A_T$ versus L the length of baseline
 for the case with standard matter effect. Left panel is for $E=5$ GeV and
 right panel is for $E=15$ GeV. Neutrino parameters are the same
 as in Fig. \ref{PVsE}.}
 \label{TsVsLA}
\end{figure}

 The analytic formula is a better approximation for shorter baseline.
 In Figs. (\ref{PVsLA}) and (\ref{TsVsLA}) we plot $P(\nu_\mu \to \nu_e)$
 versus L, the length of baseline. For $L > 10690$ km the lines
 labeled with "first order in perturbation" are computed using
 (\ref{evolSb}). Left panels in Figs. (\ref{PVsLA}) and (\ref{TsVsLA})
 are for $E=5$ GeV and right panels are for $E=15$ GeV. One can
 see clearly that the analytic result is
 a very good approximation for $L \lsim 6000$ km. It's no longer
 precise for $L \gsim 6000$ km. In all these plots the results
 computed using the perturbation theory are always in remarkable agreement
 with that of numerical computation.

\begin{figure}
\begin{flushleft}
\includegraphics[height=6.cm,width=8cm]{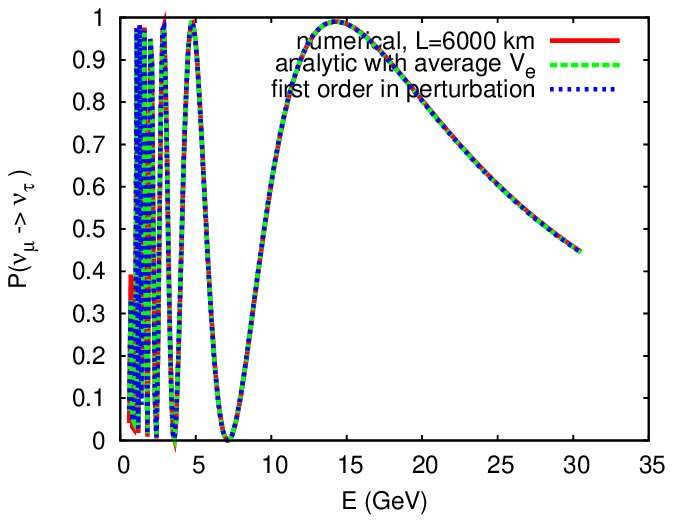}
\end{flushleft}
\begin{flushright}
\vskip -6.5cm
\includegraphics[height=6.cm,width=8cm]{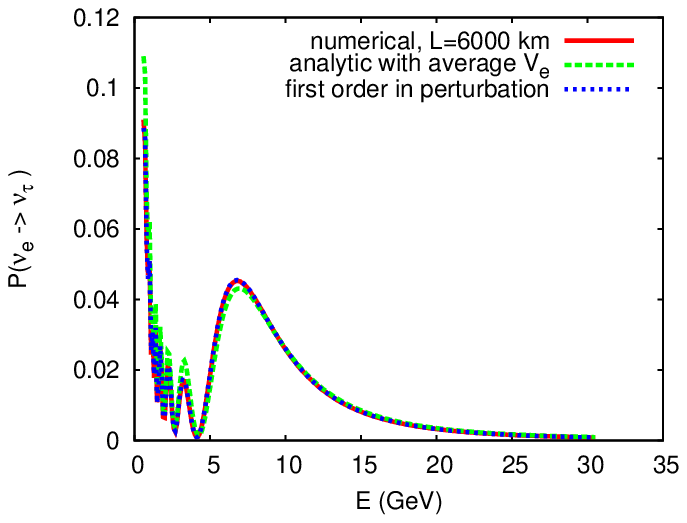}
\end{flushright}
\vskip 0.0cm \caption{\small
 Plots for the case with standard matter effect.
  Left panel, $P(\nu_\mu \to \nu_\tau)$
 versus energy, $L=6000$ km; right panel, $P(\nu_e \to \nu_\tau)$ versus energy,
 $L=6000$ km. Other parameters are given in Fig. \ref{PVsE}. }
 \label{PVsEB}
 \end{figure}

\begin{figure}
\begin{flushleft}
\includegraphics[height=6.cm,width=8cm]{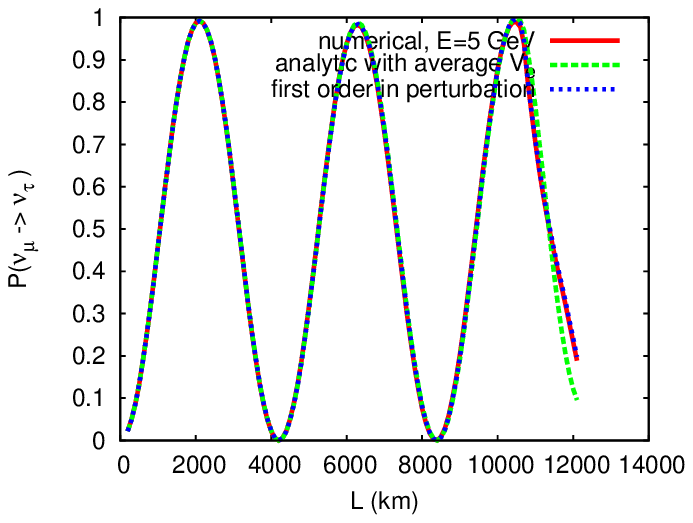}
\end{flushleft}
\begin{flushright}
\vskip -6.5cm
\includegraphics[height=6.cm,width=8cm]{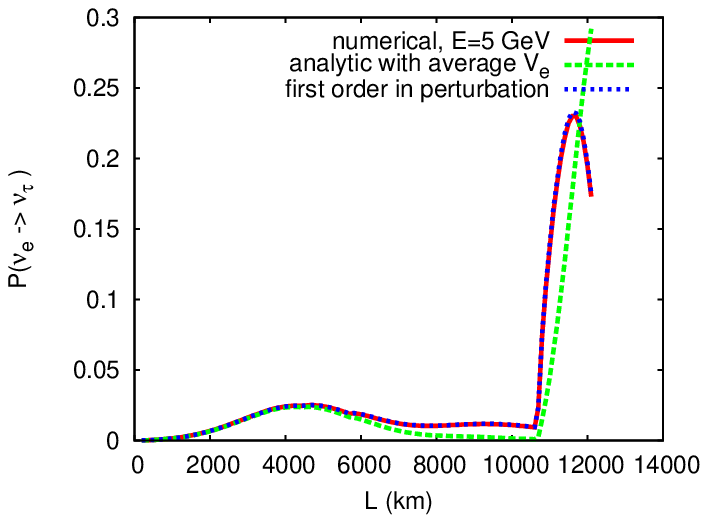}
\end{flushright}
\vskip 0.0cm \caption{\small
 Plots for the case with standard matter effect.
  Left panel, $P(\nu_\mu \to \nu_\tau)$
 versus L, $E=5$ GeV; right panel, $P(\nu_e \to \nu_\tau)$ versus L,
 $E=5$ GeV. Other parameters are given in Fig. \ref{PVsE}. }
 \label{PVsLAB}
 \end{figure}

 In Fig. \ref{PVsEB} and \ref{PVsLAB} we also compare computations on $P(\nu_\mu \to
 \nu_\tau)$ and $P(\nu_e \to \nu_\tau)$. We see that $P(\nu_\mu \to \nu_\tau)$
 is very well approximated by the analytic result computed using
 Eq. (\ref{AnalyticA}) except for $L \gsim 11500$ km.
 For $P(\nu_e \to \nu_\tau)$ there are some differences
 for $L \gsim 6000$ km. Differences of these magnitude are
 negligible in $P(\nu_\mu \to \nu_\tau)$. Again we see that the results
 of first order in the perturbation theory are in remarkable
 agreement with the results of numerical computation.

 One can understand the success of this formulation of neutrino oscillation by
 noting that indeed we are expanding in small quantities and
 we are dealing with a perturbation theory. In the following we show
 that $C_{jk}(j\neq k)$ computed using Eq. (\ref{C3}) is suppressed
 by some small quantities. Hence in $(1-i C)$ condition
 \bea
 |C_{jk}| \ll 1 ~(j \neq k) \label{cond}
 \eea
 can be satisfied in the first order. The second order result
 of order ${\cal O}(C^2)$ is further suppressed. Thus
 we can be confident on the perturbation theory. Note that
 another condition is that density changes mildly in the mantle or
 in the core, as already emphasized in section (\ref{sec2}).
 (\ref{cond}) is explained as follows. We first consider the case
 with $\Delta m^2_{31} > 0$.

 i) For $0.5 ~\textrm{GeV} \ll E < 7-10 ~\textrm{GeV}$,
 $ \frac{\Delta m^2_{21} }{2 E} \ll {\bar V}_e < \frac{\Delta m^2_{31}} {2 E}$ holds
 and this is the region below $1-3$ MSW resonance.
 In this range of energy the eigenvalues of ${\bar H}$ are
 approximately $(\frac{\Delta m^2_{21}}{2 E} \cos^2\theta_{12}, {\bar V}_e,
 \frac{\Delta m^2_{31}}{2 E})$ in the limit that correction
 of ${\cal O}(\sin\theta_{13})$ and
 ${\cal O}(\Delta m^2_{21}/(2 E {\bar V}_e) ~)$ are
 neglected. Small correction of ${\cal O}(\frac{\Delta m^2_{21}}{2 E})$
 has been neglected in larger eigenvalues. Hence the first
 entries of ${\bar H}$ are changed to the second entries. We have
 \bea
  \sin 2 \theta^m_{13}\approx \sin 2 \theta_{13},
  ~~\cos \theta^m_{12} \approx \frac{\Delta m^2_{21}}{4 E}
  \frac{1}{{\bar V}_e} \sin 2 \theta_{12}. \label{appsol1}
  \eea
 So we get $|(U_s)_{e3}|=\sin\theta^m_{13} \ll 1$ and
 $(U_s)_{e1}=\cos \theta^m_{12}\cos\theta^m_{13} \ll 1 $.
 $C_{jk} \propto (U_s)^*_{ej} (U_s)_{ek}(j\neq k)$ is suppressed either by
 $(U_s)_{e3}$ or by $(U_s)_{e1}$.

 In this range of energy $C_{jk}$ is suppressed by small quantities
 $\sin\theta_{13}$ and $\Delta m^2_{21}/(2 E {\bar V}_e)$.

 ii) For $E > 7-10$ GeV, ${\bar V}_e > \frac{\Delta m^2_{31}}{2 E}$ holds
 and this is the region above $1-3$ MSW resonance.
 In this range of energy the eigenvalues of ${\bar H}$ are
 approximately $(\frac{\Delta m^2_{21}}{2 E} \cos^2\theta_{12},\frac{\Delta m^2_{31}}{2 E},
 {\bar V}_e)$ in the limit that corrections of ${\cal O}(\sin\theta_{13})$
 and ${\cal O}(\Delta m^2_{21}/(2 E {\bar V}_e) ~)$ are neglected.
 In this case the first entries of ${\bar H}$
 are changed to the third entries. We can get (see appendix)
 \bea
 \cos\theta^m_{13}\approx \frac{\Delta m^2_{31}}{4 E} \frac{1}{{\bar V}_e}
 \sin 2 \theta_{13} \ll 1. \label{appsol2}
 \eea
 So we have $(U_s)_{e1}=\cos\theta^m_{13}\cos\theta^m_{12} \ll 1$ and
 $(U_s)_{e2}= \cos\theta^m_{13} \sin\theta^m_{12} \ll 1$. $C_{jk}$
 is suppressed either by $(U_s)_{e1}$ or by $(U_s)_{e2}$.

 In this range of energy $C_{jk}$ is suppressed by small quantity
 $\Delta m^2_{31}/(2 E {\bar V}_e) ~\sin2\theta_{13}$.

 iii) For $E \sim 7-10$ GeV, condition ${\bar V}_e = \frac{\Delta m^2_{31}}{2 E}
 \cos 2\theta_{13}$ can be satisfied and the MSW resonance
 can happen. In this case $\cos \theta^m_{12}$ is expressed
 using Eq. (\ref{appsol1}). $C_{12}$ and $C_{13}$ are
 suppressed by it. But $C_{23}$ is no longer suppressed by
 $\sin\theta^m_{13}$ or $\cos\theta^m_{13}$
 since $\theta^m_{13} \approx \pi/4$ if resonance happens.

 One can see that $C_{23}$ is suppressed by the resonance condition
 itself. In the resonance region the second and third mass
 eigenstates get almost degenerate:
 \bea
 \frac{1}{2 E}(\Delta^3_s-\Delta^2_s) \approx \frac{\Delta m^2_{31}}{2 E}\sin
 2\theta_{13}+{\cal O}(\frac{\Delta m^2_{21}}{2 E}). \label{appsol3}
 \eea
 It is clear that in the resonance region
 \bea
 \phi(x) =\frac{\Delta^3_s-\Delta^2_s}{2 E} (x-L/2) \nnb
 \eea
 is a small number. So
 \bea
 C_{23} &&=e^{-i \frac{\Delta^3_s-\Delta^2_s}{4 E} L } \int^L_0 ~dx~
 e^{-i \phi(x)} ~(U_s)^*_{e2} (U_s)_{e3} ~\delta V_e(x) \nnb \\
  && =e^{-i \frac{\Delta^3_s-\Delta^2_s}{4 E} L } \int^L_0 ~dx~
  ~(U_s)^*_{e2} (U_s)_{e3} ~\delta V_e(x)~
  [ 1- i\phi(x) -\frac{1}{2} \phi^2(x) +\cdots ]. \label{appsol4}
 \eea
 The first term in the bracket gives zero after integration. The second
 term gives zero after taking into account the fact that
  $\delta V$ is approximately symmetric in the Earth. Since major
 contribution is from $\phi^2$ term it is not hard to see that
 $|C_{23}|$ is indeed suppressed by small numbers.

 In this range of energy $C_{jk}$ is suppressed by $\Delta
 m^2_{21}/(2 E {\bar V}_e ) $ and
 $\Delta m^2_{31} L/(2 E) ~\sin2\theta_{13}$.

 If $\Delta m^2_{31} < 0$, eigenvalues of ${\bar H}$ are roughly
 $(\frac{\Delta m^2_{31}}{2 E}, \frac{\Delta m^2_{21}}{2 E} \cos^2\theta_{12}, {\bar V}_e)$.
 It is obtained by interchanging the first and third entries of ${\bar H}$.
 Similar to the discussion on point ii) we can get
 $\cos\theta^m_{13}\approx \sin2\theta_{13}
 |\Delta m^2_{31}|/(2 |\Delta m^2_{31}|+4 E {\bar V}_e)$.
 It is easy to see that we are expanding $C_{jk}$ using small quantities
 $(U_s)_{ej}^* (U_s)_{ek}(j\neq k)$ which is suppressed by $\cos \theta^m_{13}$.

 For anti-neutrino the MSW resonance happens for $\Delta m^2_{31} < 0$.
 Discussions on the perturbation theory closely follow the discussions
 for neutrinos above. Similarly, one can show that $C_{jk}(j\neq k)$
 is expanded using small quantities.
 The second order of perturbation theory is of order ${\cal O}(C^2)$.
 Hence it is further suppressed.

 We summarize that we do expansion using quantities
 $|(U_s)_{ej}^* (U_s)_{ek}| (j\neq k)$ in the perturbation theory. Except
 in the MSW resonance region these quantities are small in energy
 range $E \gsim 0.5$ GeV. In the resonance region the resonance
 condition itself guarantees the validness of the perturbative
 expansion.

\section{Oscillation with non-standard matter effect}\label{sec4}

 In this section we extend the discussion to case with non-standard
 matter effect. Previous works on non-standard matter effect
 include Refs. \cite{some2,NSI-longbase}. In this case we express the
 potential, mixing matrix, etc as
 \bea
 V(x)=V_n(x),~~{\bar V}={\bar V}_n,~~H=H_n, ~~{\bar H}={\bar H}_n,~~U_m=U_n,~~\Delta=\Delta_n .
 \label{label2}
 \eea

 Physics beyond the Standard Model can give non-standard four
 fermion interactions such as ${\bar q} \gamma_\mu q ~{\bar \nu_k}
 \gamma^\mu \nu_l$ and ${\bar e} \gamma_\mu e ~{\bar \nu_k}
 \gamma^\mu \nu_l$. Neutrino evolution in matter is modified
  by these terms. In general the potential can be written as follows
 \bea
 V_n(x)= \textrm{diag}\{V_e,0,0 \} +\begin{pmatrix}0 & V_{e\mu} & V_{e\tau} \cr
        V_{\mu e} & V_{\mu\mu} & V_{\mu \tau} \cr
        V_{\tau e} & V_{\tau \mu} & V_{\tau \tau}\end{pmatrix},
 \label{NonSHamil}
 \eea
 where $V_e=\sqrt{2} G_F N_e$ is the potential with standard charged current interaction,
 $V_{kl}$ is from non-standard NC interaction. $V_{lk}^*=V_{kl}$ because the
  Hamiltonian is hermitian.
  $x$ dependence in $V_{kl}$ has been suppressed in Eq.
 (\ref{NonSHamil}). $V_{ee}$ has been made zero in our convention. This is
 achieved by shifting the phases of neutrinos:
  $\nu_l \to e^{-i \int ~dx ~V_{ee} } \nu_l$, $V_{kl} \to V_{kl} +V_{ee}$.
  In this convention $V_{kl}$ is
 \bea
 V_{kl} &&=\sqrt{2} G_F ~[(f_{kl}-f_{ee}) N_e+(g_{kl}-g_{ee}) N_p
 +(h_{kl}-h_{ee}) N_n], \nnb \\
  &&= V_e ~[(f_{kl}-f_{ee}+g_{kl}-g_{ee})
   +(h_{kl}-h_{ee}) N_n/N_e], \label{NonSHamilA}
 \eea
 where $f_{kl}$,~$g_{kl}$ and $h_{kl}$ are the dimensionless strengths of non-standard
 four Fermion interactions
  $\sqrt{2} ~f_{kl} ~G_F ~{\bar e} \gamma_\mu e ~{\bar \nu_k} \gamma^\mu
 \nu_l$, $\sqrt{2} ~g_{kl} ~G_F ~{\bar p} \gamma_\mu p ~{\bar \nu_k} \gamma^\mu \nu_l$
 and $\sqrt{2} ~h_{kl} ~G_F ~{\bar n} \gamma_\mu n ~{\bar \nu_k} \gamma^\mu \nu_l$.
 $N_p$ and $N_n$ are number densities of proton and neutron in
 matter. The second line in Eq. (\ref{NonSHamilA}) holds because of
 the equality $N_e=N_p$ in neutral matter.

   We can re-write $V_n$ as
  \bea
  V_n=V_e \begin{pmatrix} 1 & \epsilon_{e\mu} & \epsilon_{e\tau} \cr
   \epsilon_{\mu e} & \epsilon_{\mu \mu} & \epsilon_{\mu \tau} \cr
   \epsilon_{\tau e} & \epsilon_{\tau \mu} & \epsilon_{\tau \tau}
   \end{pmatrix}. \label{NonSHamilB}
  \eea
 where $\epsilon_{kl}=V_{kl}/V_e$. As can be seen in Eq.
 (\ref{NonSHamilA}), $\epsilon_{kl}$ is not a constant if
 $N_n/N_e$ is not a constant in matter. This happens when
 chemical composition changes in matter. In the present paper
 we are not going to consider the possibility that $\epsilon_{kl}$
 changes in the neutrino trajectory. We set $\epsilon_{kl}$
 constant in the analysis. This possibility is achieved when non-standard
 interactions are from interactions of neutrino with proton or
 with electron.

      \begin{figure}
\begin{flushleft}
\includegraphics[height=6.cm,width=8cm]{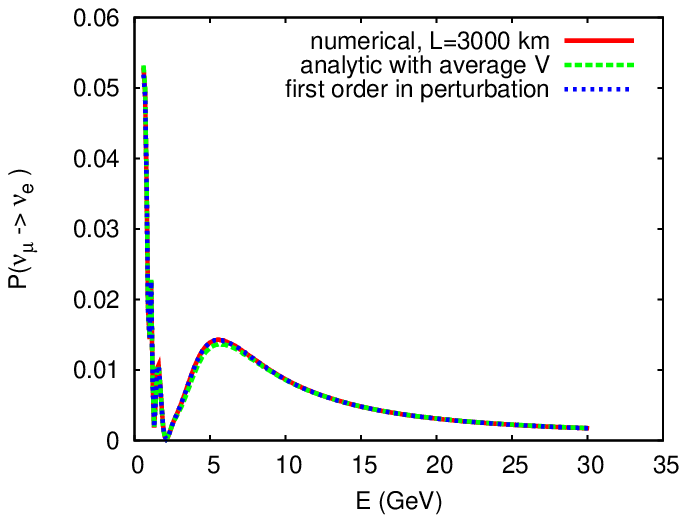}
\end{flushleft}
\begin{flushright}
\vskip -6.5cm
\includegraphics[height=6.cm,width=8cm]{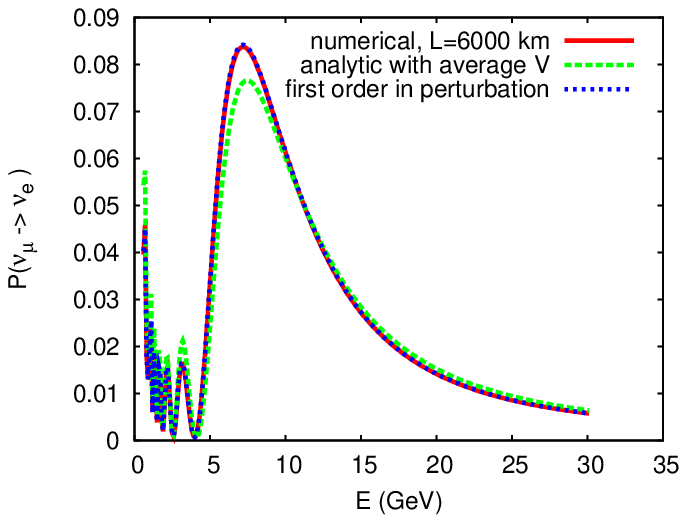}
\end{flushright}
\vskip 0.0cm \caption{\small $P(\nu_\mu \to \nu _e)$ versus energy for the
 case with non-standard matter effect.
 Left panel is for $L=3000$ km; right panel is for
 $L=6000$ km. $\epsilon_{e\mu}=0.01 ~e^{-i \pi/20}$,
  $\epsilon_{e\tau}=0.04 ~e^{-i \pi/3}$, $\epsilon_{\mu\tau}=0.01 ~e^{-i \pi/20}$.
  Other parameters are the same as in Fig. \ref{PVsE}}
 \label{PVsEC}
 \end{figure}

   \begin{figure}
\begin{flushleft}
\includegraphics[height=6.cm,width=8cm]{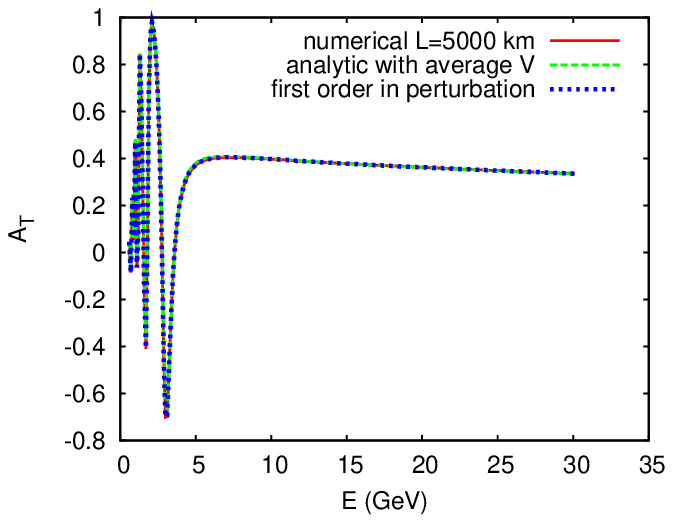}
\end{flushleft}
\begin{flushright}
\vskip -6.5cm
\includegraphics[height=6.cm,width=8cm]{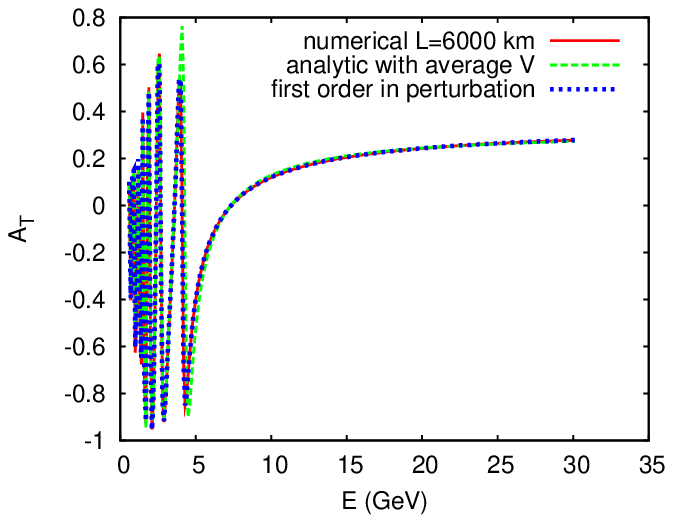}
\end{flushright}
\vskip 0.0cm \caption{\small $A_T$ versus energy for the case with
 non-standard matter effect. Left panel is for $L=3000$ km and
 right panel is for $L=6000$ km. $\epsilon_{e\tau}=0.01 ~e^{-i \pi/20}$,
  $\epsilon_{e\tau}=0.04 ~e^{-i \pi/3}$, $\epsilon_{\mu\tau}=0.01 ~e^{-i \pi/20}$.
  Other parameters are the same as in Fig. \ref{PVsE}.}
 \label{TsVsEB}
\end{figure}

 Four fermion interactions of muon neutrinos are well
 constrained by direct test, e.g. by NuTeV experiment \cite{NuTeV}.
 Since $\epsilon_{kl}$ is a coherent combination
 of strengths $f_{kl}$, ~$g_{kl}$ and $h_{kl}$, constraints on
 the strengths of these couplings from collider or fixed target experiments can
 not be directly translated to constraints on $\epsilon_{kl}$.
 However, one can have a rough constraint on the magnitude of
 $\epsilon_{kl}$. The present experiments can reach precision
 of about one percent \cite{pdg}. Hence, one can induce
 that $|\epsilon_{\mu e}|, |\epsilon_{\mu \tau}| \lsim 10^{-2}$.
 $f_{ee,e\tau,\tau\tau}$, $g_{ee,e\tau,\tau\tau}$, $h_{ee,e\tau,\tau\tau}$ are
 not well constrained because there are no powerful electron
 neutrino and tau neutrino beams.
 \footnote{Intense electron neutrino or anti-neutrino source is
 available at low energy ($E\lsim 30$ MeV) from stopped muon
 decay or from reactor. But they are not enough to make strong
 constraints.}
 Hence in our convention $\epsilon_{e\tau}$, $\epsilon_{\mu\mu}$
 and $\epsilon_{\tau \tau}$ are not well constrained by direct tests.
 Other constraints on $\epsilon_{kl}$ come from neutrino oscillation experiments.
 Previous studies\cite{constraint,constraintA} show that
 $|\epsilon_{\mu \mu}|, |\epsilon_{\tau \tau}| \lsim 10^{-2}$ and
 $|\epsilon_{e \tau}| \lsim 10^{-1}$.

 The trajectory dependent average potential and average Hamiltonian
 are
 \bea
 {\bar H}_n =H_0 + {\bar V}_n,~~
 {\bar V}_n =\begin{pmatrix}{\bar V}_e & {\bar V}_{e\mu} & {\bar V}_{e\tau} \cr
        {\bar V}_{\mu e} & {\bar V}_{\mu\mu} & {\bar V}_{\mu \tau} \cr
        {\bar V}_{\tau e} & {\bar V}_{\tau \mu} & {\bar V}_{\tau \tau}
        \end{pmatrix},
        \label{NonSHamil2}
 \eea
 where ${\bar V}_e$ is given in Eq. (\ref{aveV}). ${\bar V}_{kl}$ is
 similarly defined. Mixing matrix $U_n$ diagonalizes ${\bar H}_n$, as
 done in Eq. (\ref{defU}). $\Delta=\Delta_n$ is obtained after
 diagonalization.

 Keeping the first order result in $\delta V$ the evolution matrix is solved as
 \bea
 M_n(x) && =U_n e^{-i \frac{\Delta_n}{2 E} x} (1-i C) U_n^\dagger,
  \label{evolNS}\\
 C_{jk} && =\int^L_0 ~dx~ e^{i \frac{\Delta^j_n-\Delta^k_n}{2 E} x }
 [ (U_n)^*_{ej} ~(U_n)_{ek} ~\delta V_e(x)
 +\sum_{s,t} ~(U_n)^*_{sj} ~(U_n)_{tk} ~\delta V_{st}(x) ], \label{evolNSB}
 \eea
 where $\delta V_{st}=V_{st}-{\bar V}_{st}$. $C_{jj}=0$ holds as explained
 in section \ref{sec2}. For core crossing trajectories ($L > 10690$ km)
 the evolution matrix is generalized using Eq. (\ref{CoreMantle}) as follows
 \bea
 M_n(L)= M_{n3} ~M_{n2} ~M_{n1}. \label{evolNSR}
 \eea
 where $M_{n1,n2}$ and $M_{n3}$ are evolution matrices in the mantle and
 in the core. They are computed to the first order in $\delta V$ and using
 average potentials in the core and in the mantle separately, as explained
 in section \ref{sec2}.

 We also show result computed using analytic formula
 \bea
 M_{An}(L)= U_n ~e^{-i \frac{\Delta_n}{2 E} L} ~U_n^\dagger. \label{evolANS}
 \eea
 This formula is not generalized for core crossing trajectories.

   \begin{figure}
\begin{flushleft}
\includegraphics[height=6.cm,width=8cm]{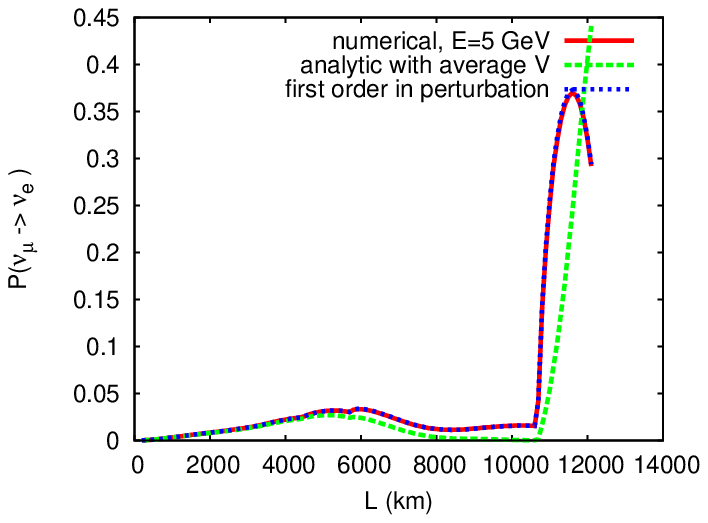}
\end{flushleft}
\begin{flushright}
\vskip -6.5cm
\includegraphics[height=6.cm,width=8cm]{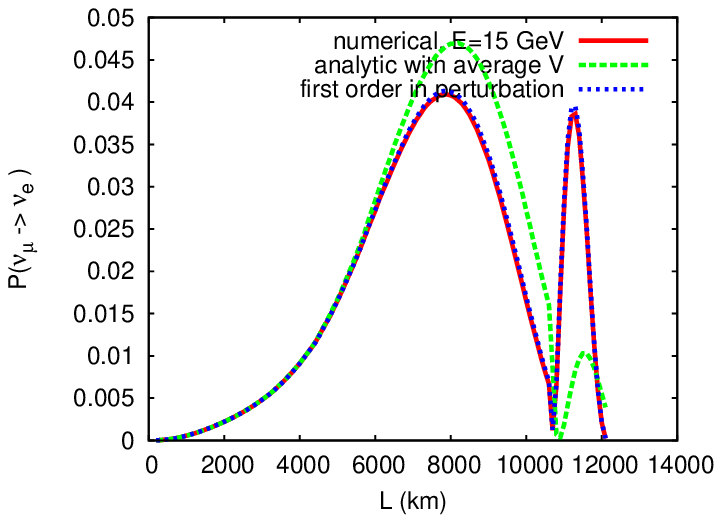}
\end{flushright}
\vskip 0.0cm \caption{\small $P(\nu_\mu \to \nu_e)$ versus L the length of
 baseline for the case with non-standard matter effect. Left panel is for $E=5$ GeV and
 right panel is for $E=15$ GeV. Neutrino parameters are the same as in Fig. \ref{PVsE}.}
 \label{PVsLB}
\end{figure}

  \begin{figure}
\begin{flushleft}
\includegraphics[height=6.cm,width=8cm]{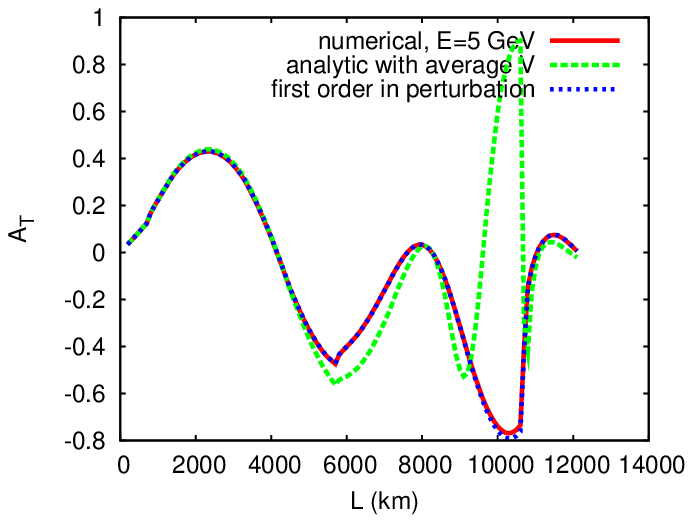}
\end{flushleft}
\begin{flushright}
\vskip -6.5cm
\includegraphics[height=6.cm,width=8cm]{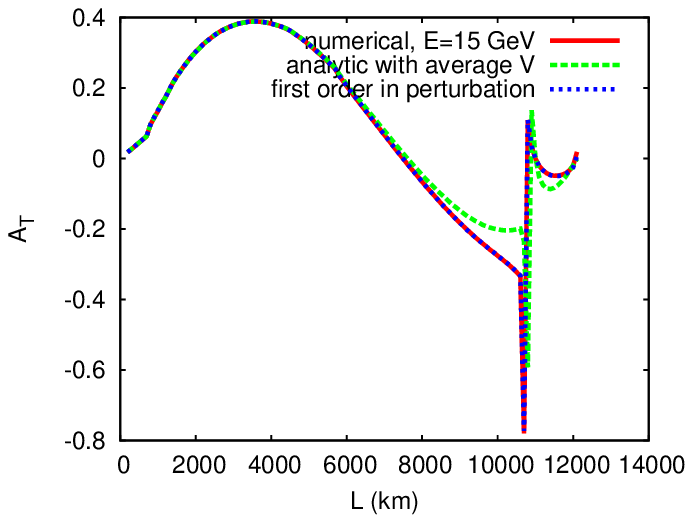}
\end{flushright}
\vskip 0.0cm \caption{\small $A_T$ versus L the length of baseline
 for the case with non-standard matter effect. Left panel is for $E=5$ GeV and
 right panel is for $E=15$ GeV. Neutrino parameters are the same
 as in Fig. \ref{PVsE}.}
 \label{TsVsLB}
\end{figure}

 In Fig. \ref{PVsEC} we plot $P(\nu_\mu \to \nu_e)$ versus energy
 for the case with non-standard matter effect. We show results of numerical
 computation, results computed using the perturbation theory and
 the results computed using the analytic formula Eq. (\ref{evolANS}).
 The left panel in the figure is for $L=3000$ km and the right panel is for
 $L=6000$ km. In the right panel one can see clearly the effect of
 MSW resonance in the first peak from right hand side.
 $\sin^2 2\theta^m_{13}$ is enhanced to about $0.1$,
 much larger than the vacuum value $0.01$ chosen in the analysis.
 The results computed in the perturbation theory are in
 perfect agreement with that of the numerical computations. The analytic
 result is not always in perfect agreement with that of the numerical
 computation. In the right panel one can see some differences of the
 two computations.
 These differences can also be seen clearly in the right panel of Fig.
 \ref{TsVsEB} where the time reversal asymmetry is plotted.

 In Fig. \ref{PVsLB} we plot $P(\nu_\mu \to \nu_e)$ versus L the length
 of the baseline. In Fig. \ref{TsVsLB} the time reversal asymmetry is
 plotted. One can see clearly the remarkable agreement between the computation
 using the perturbation theory and the numerical computation. The analytic
 computation is a good approximation for $L \lsim 6000$ km.

 We can show that indeed we are expanding in small quantities and
 we are doing a perturbation theory.
 First, the second term in the bracket of (\ref{evolNSB})
 gives small contribution. This is because $|V_{kl}|/V_e \lsim 0.1$.
 Hence $|\delta V_{kl}| L= (|\delta V_{kl}|/|V_{kl}|)
 | V_{kl}| L$ is a small number. Second, the discussion on the
 first term closely follows the discussions in the previous section.
 In lower energy region ( 0.5 GeV $\lsim E < 7-10$ GeV)
 the discussion in the previous section is still valid since $V_{kl}$ does not
 change the mixing matrix very much. For energy $E > 10$ GeV, $V_{kl}$ can be
 important. When $V_{kl}$ is important enough to determine mixing matrix
 we have (see appendix)
 \bea
 \cos\theta^m_{13} \approx \sqrt{|V_{\tau e}|^2+|V_{\mu e}|^2}/V_e \ll 1. \label{appsol5}
 \eea
 Contribution of the first term is suppressed by $(U_n)_{e1,e2}$
 which are proportional to $\cos\theta^m_{13}$.

 We summarize that in the presence of non-standard matter effect
 the perturbation theory is valid because it is expanded using
 small quantities $(U_s)^*_{ej} (U_s)_{ek}(j \neq k)$
 and $\delta V_{kl}$.

\section{Discussions and conclusions} \label{sec5}

 In this paper we propose a perturbation theory which
 simplifies the problem of neutrino oscillation in the Earth.
 We perform analysis with three flavor of neutrinos. The perturbation
 theory is developed using the trajectory dependent average potential.
 The average potential is averaged along the neutrino trajectory in
 the Earth. So it depends on the trajectory of neutrino.
 The effect of non-constant density profile in the Earth, i.e. the
 non-adiabatic contribution, is carefully included
 in the first order of the perturbation theory. This perturbation
 theory is generalized using Eq. (\ref{CoreMantle}) for core
 crossing trajectories, i.e. for $L < 10690$ km. The problem of
 neutrino oscillation is substantially simplified using the perturbation
 theory presented in the present paper.

 Using the perturbation theory we study neutrino
 oscillation in the Earth for cases with standard matter effect and with
 non-standard matter effect. It is shown that for both cases the formulation
 presented gives a precise description of the neutrino oscillation.
 We study observables of flavor conversion and of time reversal asymmetry in neutrino
 oscillation. We find remarkable agreement between result of computation
 using our perturbation theory and that of numerical computation.
 This is a nontrivial agreement. To see that it is nontrivial one can try
 another perturbation theory. For the case with non-standard matter effect
 one may rewrite the hamiltonian as follows
 \bea
 H= H_1+H_2, ~~ H_1=H_0+\textrm{diag}\{ {\bar V}_e, 0,0\},
 ~~H_2=\begin{pmatrix} \delta V_e & V_{e\mu} & V_{e\tau} \cr
        V_{\mu e} & V_{\mu\mu} & V_{\mu \tau} \cr
        V_{\tau e} & V_{\tau \mu} & V_{\tau \tau}
        \end{pmatrix}. \nnb
 \eea
 One may first solve evolution problem of $H_1$ and treat $H_2$ as
 perturbation. The first order result of this perturbation theory is
 a good approximation for $E \lsim 5$ GeV. Apparently it will not be
 good for higher energy neutrinos when non-standard matter effect
 becomes important. It is not able to reproduce the time reversal asymmetry
 in high energy region which is dominated by non-standard matter effect.
 To improve the perturbation theory one may go to higher order
 and make the description more complicated. Apparently this perturbation
 theory is not a good choice compared to one given in the present
 paper.

 We also show the results of analytic
 computation, i.e. the computation using average potential
 but without non-adiabatic contribution. It is shown that the
 analytic result gives a very good approximation to the oscillation
 for $L \lsim 6000$ km.
 It means that the problem of neutrino oscillation can be greatly simplified
 for $L \lsim 6000$ km and $E \gsim 0.5$ GeV. For this range of
 parameter space matter effect in a fixed baseline is described by
 a constant potential term. For standard matter effect, in particular, it
 says that matter effect is a one parameter fit.

 The perturbation theory given in the present paper is valid for $E \gsim 0.5$ GeV.
 Previous works \cite{hls, other} cover the energy range $E \lsim 30$ MeV.
 They are all energy ranges away from the $1-2$ MSW resonance. It is a
 question to find a compact and simple formulation of neutrino oscillation
 in the Earth in the energy range $30 \textrm{~MeV} \lsim E \lsim 500$ MeV
 for which the $1-2$ MSW resonance can happen in the Earth.

 We perform analysis using the PREM density profile \cite{PREM}.
 It is a symmetric density profile. We also assume $\epsilon_{kl}$
 is constant in considering non-standard matter effect. These assumptions are
 not generally true. For more general density profile further research
 is needed to understand whether the perturbation theory is
 as good as analyzed in the present paper. We expect that this perturbation
 theory is still a good approximation in more general cases. Research on
 neutrino evolution with more general density profile will be presented
 in further publications.

{\bf Acknowledgment:} The research is supported in part by National
Science Foundation of China(NSFC), grant 10745003.

\section*{\bf Appendix}
 In this appendix we show that in large $V_e$ limit $\cos\theta^m_{13}$ is
 small. We write
 \bea
 H=H^x+\textrm{diag}\{V_e,0,0\}, \label{A1}
 \eea
 where
 \bea
 H^x=H_0, \label{A2}
 \eea
 for the case with standard matter effect, and
 \bea
 H^x=H_0+\begin{pmatrix} 0 & V_{e\mu} & V_{e\tau} \cr
        V_{\mu e} & V_{\mu\mu} & V_{\mu \tau} \cr
        V_{\tau e} & V_{\tau \mu} & V_{\tau \tau}
        \end{pmatrix}, \label{A3}
 \eea
 for the case with non-standard matter effect.

 We consider the eigenvalue problem of $H$:
 \bea
 H ~X = \lambda ~X, ~~X=(x_1,x_2,x_3)^T. \label{A4}
 \eea
 In the large $V_e$ limit one of the engenvalues of $H$
 is roughly $V_e$. So at zeroth order $\lambda \approx V_e$.
 We solve the eigenvalue problem using perturbation in $H^x/V_e$:
 \bea
 \lambda=\lambda_0+\lambda_1+\lambda_2+\cdots , ~~X=X_0 +X_1+X_2 +\cdots \label{A5}
 \eea
 So to first order in $H_x/V_e$ we get
 \bea
 & \textrm{diag}\{V_e,0,0\} ~X_0 =\lambda_0 ~X_0 , \label{A6} \\
 & H^x ~X_0+ \textrm{diag}\{V_e,0,0\} ~X_1 = \lambda_0 ~ X_1+\lambda_1 X_0. \label{A7}
 \eea
 We get from Eq. (\ref{A6})
 \bea
 \lambda_0 =V_e, ~~X_0= (1,0,0)^T .\label{A8}
 \eea
 Putting Eq. (\ref{A8}) into Eq. (\ref{A7}) we get
 \bea
 \lambda_1=H^x_{ee}, ~~X_1=(0, H^x_{\mu e}/V_e, H^x_{\tau e}/V_e )^T. \label{A9}
 \eea

 Writing $X=(\sin \theta^m_{13} e^{-i \delta^m_{13} }, \cos\theta^m_{13} \sin \theta^m_{23},
 \cos \theta^m_{13} \cos \theta^m_{23} )^T$, we can get
 \bea
 \cos^2\theta^m_{13} \approx (|H^x_{\mu e}|^2+|H^x_{\tau e}|^2)/V^2_e. \label{A10}
 \eea
 Using Eq. (\ref{evol1b}) in the case with standard matter effect we can get
 \bea
 \cos\theta^m_{13}= \frac{|\Delta m^2_{31}|}{2 E} \sqrt{1-|(U_0)_{e3}|^2} ~|(U_0)_{e3}|
  =\frac{|\Delta m^2_{31}|}{4 E V_e} sin2\theta_{13}. \label{A11}
 \eea
 In the case with non-standard matter effect there are two possibilities.
 One possibility is that effect of $H_0$ is larger than the non-standard matter effect.
 Hence, Eq. (\ref{A11}) holds and we have small $\cos\theta^m_{13}$ in large
 $V_e$ limit. The other possibility is that non-standard matter effect is large enough
 to determine the mixing matrix. So we have
 \bea
 \cos \theta^m_{13} = \sqrt{|V_{\mu e}|^2 +|V_{\tau e}|^2 }/V_e.
 \label{A12}
 \eea
 We conclude that $\cos\theta^m_{13}$ is
 a small quantity in large $E$ limit for both the case with standard matter effect
 and the case with non-standard matter effect.

\end{document}